\documentclass[aps,prd,twocolumn,superscriptaddress,runinaddress,floatfix,preprintnumbers,showpacs]{revtex4}
\usepackage{graphicx}
\usepackage{dcolumn}

\usepackage{amsmath}
\usepackage{amssymb}
\usepackage{amsfonts}
\usepackage{amsthm}

\newcommand{\del}{\partial}

\newcommand{\nn}{\nonumber}

\newtheorem{theorem}{Theorem}[section]
\newtheorem{corollary}[theorem]{Corollary}
\newtheorem{lemma}{Lemma}[section]
\newtheorem{definition}{Definition}[section]
\newtheorem{proposition}{Proposition}[section]
\newtheorem{example}{Example}[section]

\newcommand{\eqn}[1]{ \begin{equation} #1 \end{equation} }

\def\be{\begin{eqnarray}}
\def\ee{\end{eqnarray}}

\begin{document}

\preprint{ADP-10-20-T716}

\title{Wave Functions of the Proton Ground State in the Presence of a Uniform Background Magnetic Field in Lattice QCD}


\author{Dale S. Roberts}
\affiliation{Special Research Centre for the Subatomic Structure of Matter and Department of Physics, University of Adelaide 5005, Australia. }
\author{Patrick O. Bowman}
\affiliation{Centre for Theoretical Chemistry and Physics and Institute of Natural Sciences, Massey University (Albany), Private Bag 102904, North Shore City 0745, New Zealand}
\author{Waseem Kamleh}
\author{Derek B. Leinweber}
\affiliation{Special Research Centre for the Subatomic Structure of Matter and Department of Physics, University of Adelaide 5005, Australia. }

\begin{abstract}
We calculate the probability distributions of quarks in the ground
state of the proton, and how they are affected in the presence of a
constant background magnetic field. We focus on wave functions in the
Landau and Coulomb gauges. We observe the formation of a scalar
$u$-$d$ diquark clustering. The overall distortion of the quark
probability distribution under a very large magnetic field, as
demanded by the quantisation conditions on the field, is quite
small. The effect is to elongate the distributions along the external
field axis while localizing the remainder of the distribution.

\end{abstract}

\maketitle

\section{Introduction}

The wave function of a baryon on the lattice provides insight into
the shape and properties of the particle. Furthermore, the wave
function also provides a diagnostic tool for the lattice, being able
to determine how well a particular state fits on the lattice
volume. The earliest work on wave functions on the lattice was carried
out on small lattices, for the pion and rho, initially in
$SU(2)$ \cite{Velikson:1984qw}. Further progress was made in the early
nineties, where gauge invariant Bethe-Salpeter amplitudes were
constructed for the pion and rho \cite{Chu:1990ps,Gupta:1993vp} by choosing a path
ordered set of links between the quarks. This was then used to
qualitatively show Lorentz contraction in a moving pion. Hecht and
DeGrand \cite{Hecht:1992uq} conducted an investigation on the wave
functions of the pion, rho, nucleon and Delta using a gauge dependent
form of the Bethe-Salpeter amplitude, primarily focusing on the
Coulomb gauge.

The background field method \cite{Smit:1986fn} for placing an external
electromagnetic field on the lattice has been used extensively in
lattice QCD to determine the magnetic moments of hadrons. Early studies on
very small lattices with only a few
configurations \cite{Martinelli:1982cb,Bernard:1982yu} showed
remarkable agreement with the experimental values of the magnetic
moments of the proton and neutron. More recent studies on magnetic
moments \cite{Lee:2005ds} have shown good agreement with experimental
values of the magnetic moments of the baryon octet and decuplet. This
method has also been extended to the calculation of magnetic and
electric polarisabilities \cite{Burkardt:1996vb,Lee:2005dq}. Here we use the
wave function to determine the effect of the background magnetic
fields on the shape of the proton.

As background field methods have become more widely used, it is
apparent that the large fields demanded by the quantisation conditions
should cause some concern with regards to the calculation of moments
and polarisabilities. It is entirely possible that the distortion
caused by these fields could be so dramatic that the particle under
investigation bears little resemblance to its zero-field form. For this reason, we will use the
wave function as a tool to investigate the deformation caused by a
background field on a particle.


\section{Wave Function Operators} \label{WfunOps}
The wave function of a baryon on the lattice is defined to be
proportional to the two-point correlation function at zero momentum in
position space. The two-point correlation function in position space
for a proton can be written as \be G(\vec{x},t) =
\langle\Omega\vert T\{\chi_P(x)\bar{\chi}_P(0)\}\vert\Omega\rangle,
\label{twoptdef}
\ee where the Dirac indices have been suppressed. The operators
$\bar{\chi}_P$ and $\chi_P$ create and annihilate the proton
respectively. $\chi_P$ is given by
\eqn{\chi_P(\vec{x}) = \epsilon^{abc}(u^T_a(\vec{x})C\gamma_5d_b(\vec{x}))u_c(\vec{x}),}
where $u$ and $d$ are the Dirac spinors for the up quark and down
quark respectively and $C = \gamma_2\gamma_4$ is the charge
conjugation matrix in the Pauli representation, with Dirac indices
suppressed and colour indices present. This interpolating field is
chosen as it couples strongly to the ground state of the proton. From
this, we construct the adjoint spinor that will create the proton:
\begin{align}
\bar\chi_P(\vec{x}) &= \chi^{\dagger}_P\gamma_0\nn \\
&= \epsilon^{abc}\bar{u}_a(\vec{x})(\bar{d}_b(\vec{x})C\gamma_5\bar{u}^T_c(\vec{x})).
\end{align}
In order to construct the wave function across the entire lattice, we
need to modify the definition of the annihilation operator to be able
to annihilate each of the quarks at different points on the lattice
with respect to some central point or origin. In this case, we wish to have
two quarks annihilate some distance in one dimension from $\vec{x}$ and have
 the remaining quark annihilate
at any other point on the lattice with respect to $\vec{x}$. This
gives
\begin{align}
\chi_P(\vec{x},\vec{y},\vec{z},\vec{w}) &= \epsilon^{abc}(u^T_a(\vec{x}+\vec{y})C\gamma_5d_b(\vec{x}+\vec{z}))u_c(\vec{x}+\vec{w}).
\label{interpfielddef}
\end{align}
For the case of a $d$ quark wave function, we select $\vec{w} =
(d_1,0,0)$ and $\vec{y} = (d_2,0,0)$. For separations of the $u$
quarks across even numbers of lattice sites, $d_1 = -d_2$, and for odd
separations, $d_1+1 = -d_2$.  We consider eight values for the
separation of the quarks in Eq (\ref{bigwfundef}), between 0 and 7
lattice spacings, or $0\, \mathrm{fm}$ to $0.896\, \mathrm{fm}$. Wave
functions of the $u$ quarks are explored in a similar manner. By
inserting this into the definition of the two-point correlation
function in Eq.~(\ref{twoptdef}) and restoring Dirac indices, we
arrive at the definition of the wave function operator
\begin{widetext}
\begin{align}
G_{\gamma\delta}(\vec{x},\vec{y},\vec{z},\vec{w},t) =& \epsilon^{abc}\epsilon^{a^\prime b^\prime c^\prime } (C\gamma_5)_{\alpha\beta}(C\gamma_5)_{\mu\nu}\langle\Omega\vert u^a_\alpha(\vec{x}+\vec{y})d^{\,b}_\beta(\vec{x}+\vec{z})u^c_\gamma(\vec{x}+\vec{w})\bar{u}^{a^\prime}_\delta(0)\bar{d}^{\,b^\prime}_\mu(0)\bar{u}^{c^\prime}_\nu(0)\vert\Omega\rangle\nn \\
=&-\epsilon^{abc}\epsilon^{a^\prime b^\prime c^\prime } S^{cc^\prime}_{u\gamma\delta}(\vec{x}+\vec{w},0)\bigl( \mathrm{Tr}(S^{aa^\prime}_u(\vec{x}+\vec{y},0)(C\gamma_5S^{bb^\prime}_d(\vec{x}+\vec{z},0)C\gamma_5)^T \nn \\ &+(C\gamma_5S^{bb^\prime}_d(\vec{x}+\vec{z},0)C\gamma_5)_{\gamma\alpha}^TS^{aa^\prime}_{u\alpha\delta}(\vec{x}+\vec{y},0)_{\gamma\delta}\bigr),
\label{bigwfundef}
\end{align}
\end{widetext}
where the required Wick contractions have been taken over the quark
spinors, and $S_u(\vec{x},0)$ and $S_d(\vec{x},0)$ represent
propagators for the $u$ and $d$ quarks respectively propagating from
$0$ to $\vec{x}$. A sum over $\vec{x}$ is used to isolate the zero
momentum state. Note that this definition of the wave function is not
gauge invariant, and as such, gauge fixing is required. For large Euclidean times, 
\eqn{ \sum_{\vec{x}} G_{\gamma\delta}(\vec{x},\vec{y},\vec{z},\vec{w},t)=\lambda_0\lambda(\vec{y},\vec{z},\vec{w})e^{-Mt}\Bigl( \frac{1+\gamma_0}{2}\Bigr)_{\gamma\delta}, \label{wfunLargetDef} }

where $\lambda_0$ is the coupling of the source interpolator to the
ground state of mass $M$ (or energy $E$ in the external field case)
and $\lambda(\vec{y},\vec{z},\vec{w})$ encapsulates information on the
ground state wave function. Thus, $G$ is directly proportional to
the wave function. Through our use of gauge invariant Gaussian
smearing at the source, the standard two point function as in Eq.~(\ref{twoptdef}) and the wave
function at the source are gauge invariant.
\section{Simulation Details} \label{SimDetails}

As this is the first investigation of the effects of a magnetic field
on the wave function of the nucleon, we use an ensemble of 200
quenched configurations with a lattice volume of $16^3\times32$,
generated using the Luscher-Weisz $\mathcal{O}(a^2)$ improved gauge
action \cite{Luscher:1984xn}. The $\mathcal{O}(a)$ improved FLIC
(Fat-Link Irrelevant Clover) fermion action \cite{Zanotti:2004qn} is
used to generate the quark propagators with fixed boundary conditions
in the time direction. Four sweeps of stout link smearing
\cite{Morningstar:2003gk} with smearing parameter $\rho = 0.1$ are
applied to the gauge links in the irrelevant operators of the FLIC
action. We use $\beta=4.53$, corresponding to a lattice spacing of
$a=0.128 \,\mathrm{fm}$, determined by the Sommer parameter, $r_0 =
0.49 \,\mathrm{fm}$ \cite{Sommer:1993ce}. We employ 50 sweeps of gauge
invariant Gaussian smearing \cite{Gusken:1989qx} to the fermion source
at time slice 8. Two values for the hopping parameter are considered,
$\kappa=0.12885$ and $0.12990$, corresponding to pion masses of
$0.697\,\mathrm{GeV}$ and $0.532\,\mathrm{GeV}$. The gauge fields
generated are fixed to the Landau gauge using the conjugate gradient
Fourier acceleration method for improved actions \cite{Davies:1987vs},
to an accuracy of 1 part in $10^{12}$.

The normalisation chosen for the wave function is to scale the raw
correlation function data such that the sum (over $\vec{x}$ and the
parameter associated with the quark wave function coordinate) of the
square of the correlation function is $1$ for each Euclidean time,
$t$. For the $d$ quark, this is given by
\eqn{\xi^2(t)\frac{1}{V}\sum_{\vec{z},\vec{x}}G_{\gamma\delta}^\star(\vec{x},0,\vec{z},0,t)G_{\gamma\delta}(\vec{x},0,\vec{z},0,t)=1, \label{cfuncnorm}}
and similarly for the $u$ quarks, with no sum over $\gamma$ or
$\delta$. Here, $V$ is the spatial volume of the lattice. Note that the
quark separation parameters $d_1$ and $d_2$ are zero here. The wave
functions of other quark separations are then scaled by the same
factor, $\xi(t)$. In reporting our results, we focus on the
probability distribution,
\eqn{\rho_{\gamma\delta}=\xi^2(t)\frac{1}{V}\sum_{\vec{x}}G_{\gamma\delta}^\star(\vec{x},\vec{y},\vec{z},\vec{w},t)G_{\gamma\delta}(\vec{x},\vec{y},\vec{z},\vec{w},t).
\label{probdistdef}}
 For the zero field case, we report the probability
distribution from the average of spin-up, $(\gamma,\delta)=(1,1)$ and spin
down, $(\gamma,\delta)=(2,2)$ correlators. For finite $\vec{B}$, spin up
and spin down probability distributions are reported individually.  The time $t$
is selected to lie well within the ground state dominant regime as
identified by a standard covariance-matrix analysis of the local
two-point function.

\section{Zero-Field Results} \label{ZFResults}

\begin{figure*}[tph]
  \begin{center}
   $\begin{array}{c@{\hspace{0.3cm}}c}  
  \includegraphics[width=0.49\linewidth]{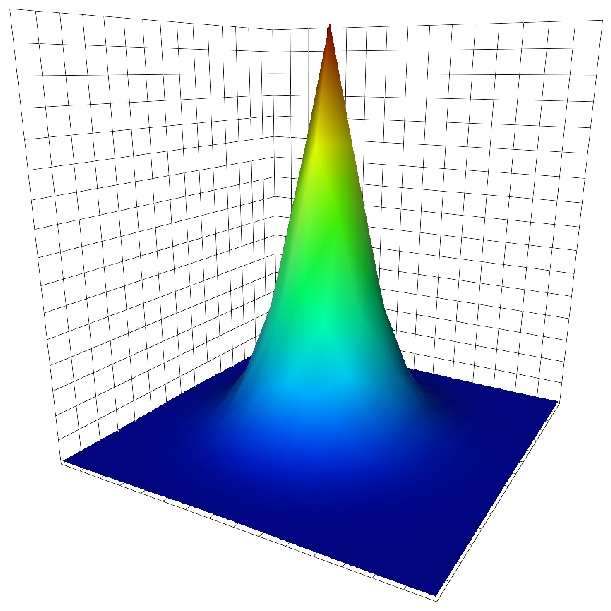} &
  \includegraphics[width=0.49\textwidth]{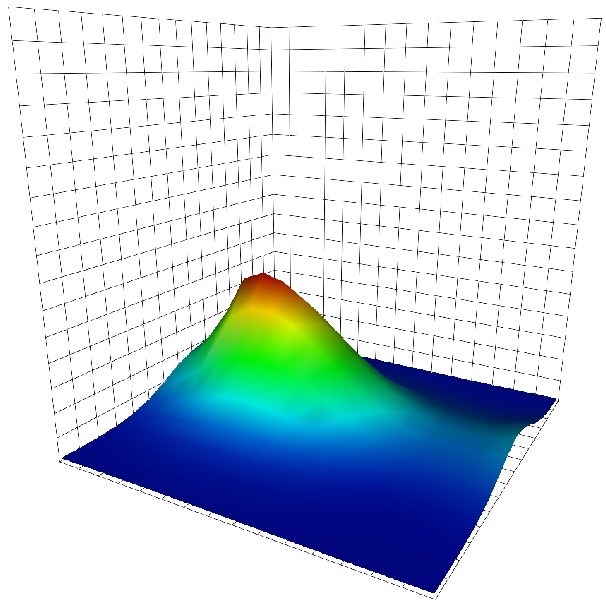} 
    \end{array}$
    \caption{(Colour Online) The Landau gauge probability distribution
      for the $d$ quark of the proton from Eqs.~(\ref{bigwfundef}) and (\ref{probdistdef}), in the plane of
      the $u$ quarks separated by zero lattice units (left), and by
      7 lattice units (right). The $d$ quark is seen to prefer to
      reside near the $u$ quark which is placed in the scalar pair in
      the interpolating field of Eq.~(\ref{interpfielddef}). }
   \label{UnsymLandauSurfPlot}
  \end{center}
\end{figure*}

We begin by looking at the probability distribution of the $d$ quark
with the aforementioned $u$ quark separations in the Landau
gauge. Immediately we notice that the probability distribution is not
symmetric around the centre of mass of the proton. We note that in
Fig. \ref{UnsymLandauSurfPlot}, the peak is centred around the $u$
quark that resides in the scalar pairing with the $d$ quark in
Eq.~(\ref{interpfielddef}). This leads us to believe that the $u$ and
$d$ quarks tend to form a scalar pair within the proton. At this
point, we choose to anti-symmetrise the idnetical $u$ quarks, changing
our annihilation operator from Eq. (\ref{interpfielddef}) to
\begin{align}
\chi_P(\vec{x},\vec{y},\vec{z},\vec{w}) &= \epsilon^{abc}(u^T_a(\vec{x}+\vec{y})C\gamma_5d_b(\vec{x}+\vec{z}))u_c(\vec{x}+\vec{w})\nn \\
&+ \epsilon^{abc}(u^T_a(\vec{x}+\vec{w})C\gamma_5d_b(\vec{x}+\vec{z}))u_c(\vec{x}+\vec{y}).
\label{symInterpField}
\end{align}
This choice is motivated by the fact that the interpolating field
places one of the $u$ quarks permanently within the scalar pair,
however, physically, this would not be the case, as the $u$ quarks
within the proton should be indistinguishable.

\begin{figure*}[tph]
  \begin{center}
   $\begin{array}{c@{\hspace{0.3cm}}c}  
  \includegraphics[width=0.49\linewidth]{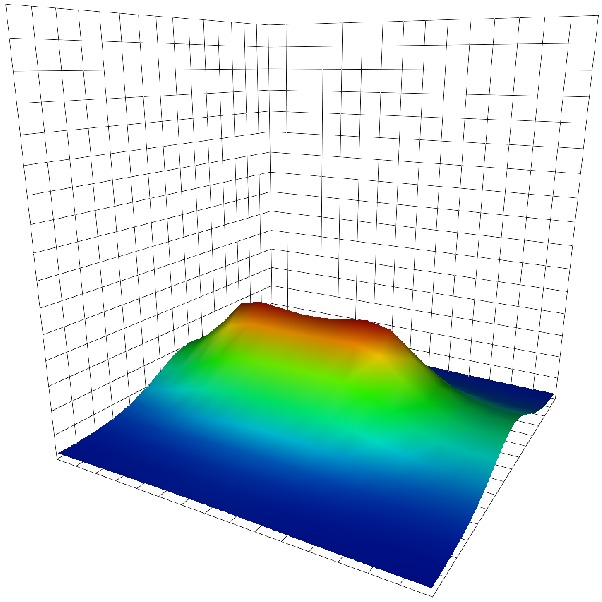} &
  \includegraphics[width=0.49\textwidth]{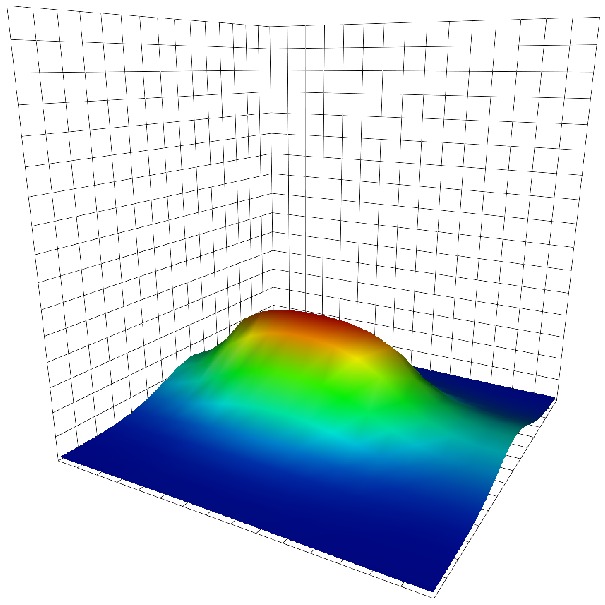}
    \end{array}$
    \caption{(Colour Online) The probability distribution for the $d$
      quark of the proton in the plane of the $u$ quarks separated
      by 7 lattice units, in the Landau gauge (left), and the Coulomb
      gauge (right). Two distinct peaks have formed over the location
      of the $u$ quarks in the Landau gauge probability distribution, whereas a
      single, broad peak is visible over the centre of mass of the
      system in the Coulomb gauge. Note: as discussed following
      Eq.~(\ref{cfuncnorm}) the scale is such that the largest value
      of all of the fixed quark separations will sit at the top of the
      grid, with all other points of the probability distribution scaled
      accordingly.}
   \label{LandauVCoulombd7SurfPlot}
  \end{center}
\end{figure*}

Upon implementing this symmetrisation, we see no evidence that diquark
clustering is occurring at small $u$-quark separations. Rather, the
probability distribution broadens and flattens around the centre of
mass of the system. However, when we move to a separation of five or
more lattice units, or $0.640\,\mathrm{fm}$, we see the formation of
two distinct peaks as illustrated in Fig.~\ref{LandauVCoulombd7SurfPlot}. At
this stage, the $u$ quarks are separated further than was considered
in \cite{Hecht:1992uq}.

\begin{figure*}[tph]
  \begin{center}
    \includegraphics[width=0.45\linewidth,angle=90]{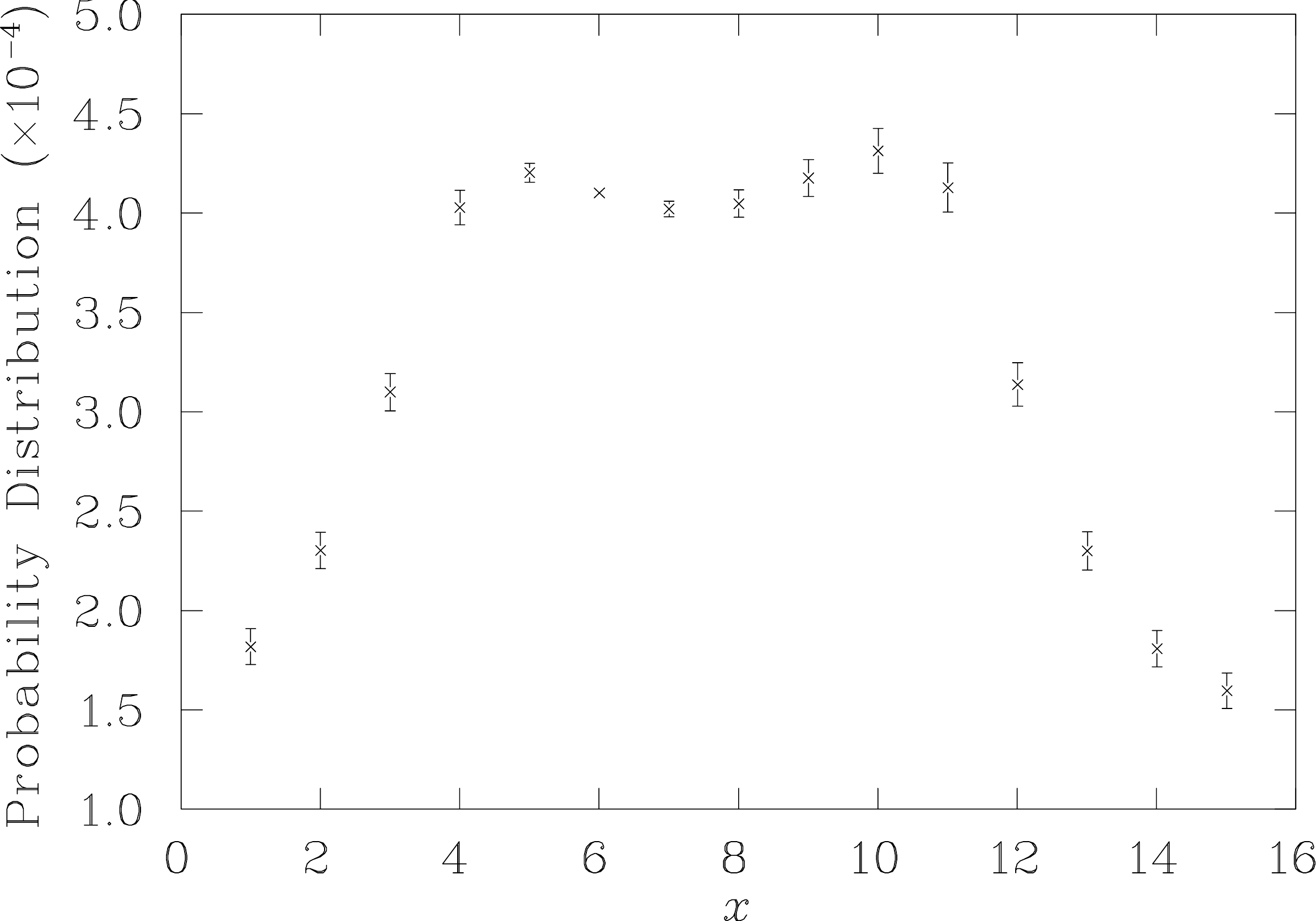}
    \caption{ The probability distribution of the $d$ quark in the
      proton with the $u$ quarks 7 lattice units apart along the $x$
      axis at $x=4$ and $11$. To clearly display the double
      peak structure, uncertainties are reported relative to the
      distribution at $x=6$.  }
    \label{DQerrPlot}
  \end{center}
\end{figure*}

  To more clearly illustrate this double peaked structure, we plot
  values of the probability distribution along the line joining the
  two fixed quarks in Fig.~\ref{DQerrPlot}. We have taken advantage of
  correlations in the uncertainties in the lattice results and present
  the uncertainty relative to the value at $x=6$.

\begin{figure*}[tph]
  \begin{center}
   $\begin{array}{c@{\hspace{0.3cm}}c}  
  \includegraphics[width=0.49\linewidth]{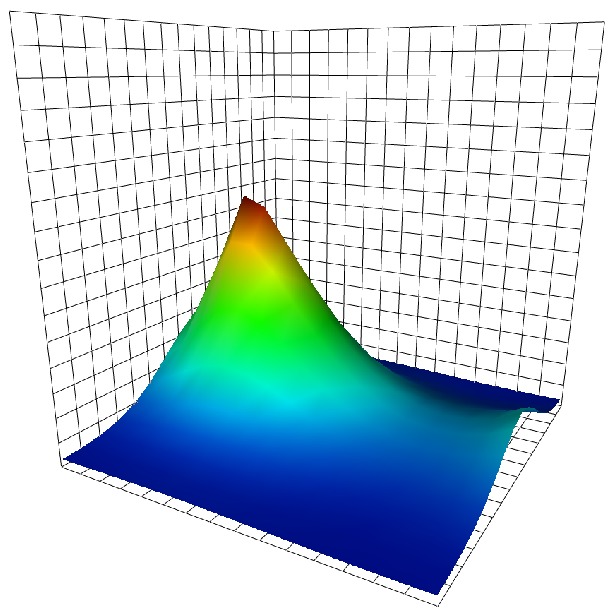} &
  \includegraphics[width=0.49\textwidth]{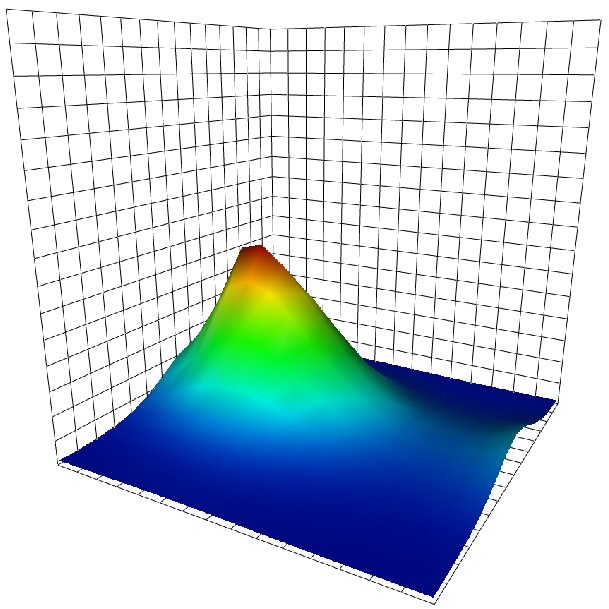} 
    \end{array}$
    \caption{(Colour Online) The probability distribution for the scalar $u$
      quark of the proton in the plane of the $u$ and $d$ quarks
      separated by 7 lattice units, in the Landau gauge (left),
      and the Coulomb gauge (right). In both gauges, the $u$ quark is
      seen to prefer to be nearer the $d$ quark. However, in the
      Coulomb gauge, the scalar $u$ quark is closer to the centre of
      the lattice than in the Landau gauge probability distribution. The scale is
      as described in Fig.~\ref{LandauVCoulombd7SurfPlot}}
   \label{LandauVCoulombd6SUSurfPlot}
  \end{center}

  \begin{center}
   $\begin{array}{c@{\hspace{0.3cm}}c}  
  \includegraphics[width=0.49\linewidth]{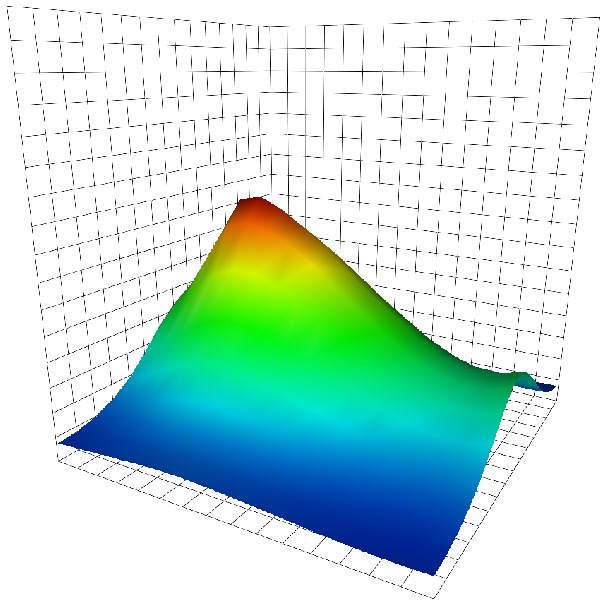} &
  \includegraphics[width=0.49\textwidth]{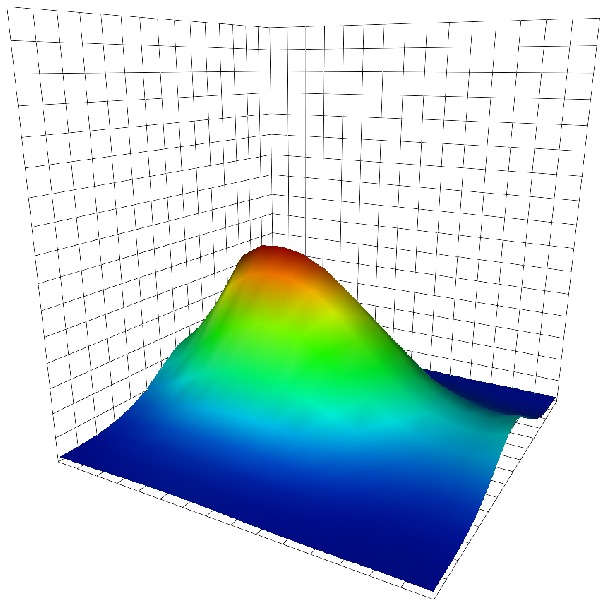}
    \end{array}$
    \caption{(Colour Online) The probability distribution for the vector $u$
      quark of the proton in the plane of the $u$ and $d$ quarks
      separated by 7 lattice units, in the Landau gauge (left),
      and the Coulomb gauge (right). The probability distribution is
      similar to the $d$ quark probability distribution in that strong
      clustering is seen in the Landau gauge. The Coulomb gauge
      results here reveal a small amount of preferred clustering with
      the $d$ quark. Also of note is that these probability distributions show
      less structure than the others, as can be seen by the height of
      the smallest values, with the scale as described in
      Fig.~\ref{LandauVCoulombd7SurfPlot}}
   \label{LandauVCoulombd6VUSurfPlot}
  \end{center}
\end{figure*}

In the Coulomb gauge, diquark clustering is present as evidenced in
the unsymmetrised wave function, however, the support in the
centralized region hides the diquark clustering upon
symmetrisation. Figure \ref{LandauVCoulombd7SurfPlot} illustrates
results for $u$ quarks separated by 7 lattice units. Such a difference
in the probability distribution between the two gauges is a remarkable
result.

In both the Landau and Coulomb gauges, the mass dependence of the
probability distributions is almost negligible, as there are no
significant differences in the shape of the probability distribution
when the quark mass is changed. This was also noted in Refs.~\cite{Velikson:1984qw,Chu:1990ps}

When we look at the probability distribution of the scalar $u$ quark (i.e. the
$u$ quark in the scalar pair with the $d$ quark in
Eq. \ref{interpfielddef}) diquark clustering becomes more pronounced in
the Landau gauge, as well as becoming apparent in the Coulomb gauge as illustrated in Fig.~\ref{LandauVCoulombd6SUSurfPlot}.

The probability distribution of the vector $u$ quark (i.e. the $u$ quark that
carries the spinor index of $\chi_P$ in Eq.~(\ref{interpfielddef})) in
the Landau gauge also exhibits diquark clustering without a direct
spin correlation in the interpolating field. Such a clustering is
anticipated in constituent quark models with hyperfine interactions.
Clustering is also observed in the Coulomb gauge. However, much like
the $d$ quark, the probability distribution is more towards the centre of mass of
the system (Fig \ref{LandauVCoulombd6VUSurfPlot}).

\begin{figure*}[tph]
  \begin{center}
   $\begin{array}{c@{\hspace{0.3cm}}c}  
  \includegraphics[width=0.49\linewidth]{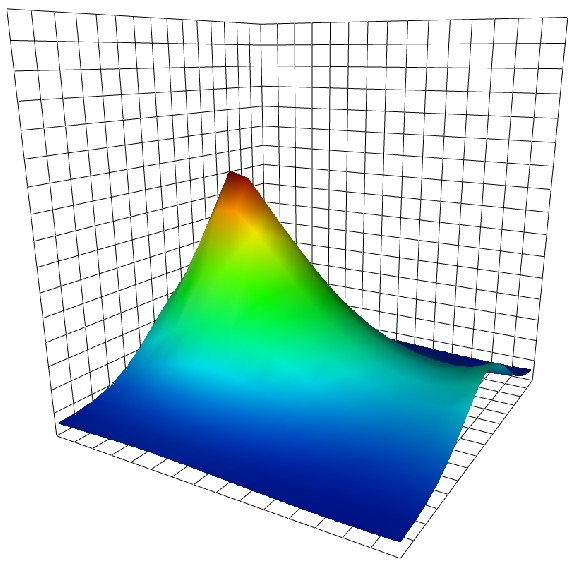} &
  \includegraphics[width=0.49\textwidth]{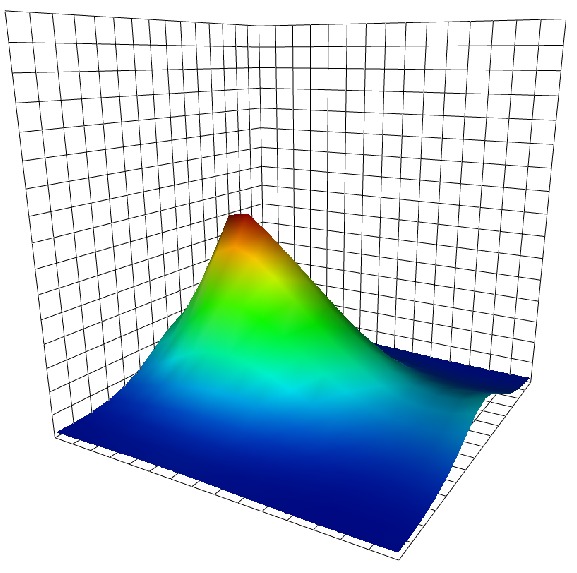}
    \end{array}$
    \caption{  (Colour Online) The probability distribution for an
      anti-symmetrised $u$ quark of the proton in the plane of the
      remaining quarks which are separated by 7 lattice units, in the Landau
      gauge (left), and the Coulomb gauge (right). In contrast to the
      $d$ quark probability distribution, a single peak is visible
      above the location of the $d$ quark in both the Coulomb and the
      Landau gauge. Note: as discussed following Eq.~(7) the scale is
      such that the largest value of all of the fixed quark
      separations will sit at the top of the grid, with all other
      points of the probability distribution scaled accordingly.} 
   \label{LandauVCoulombUQSymd7}
  \end{center}
\end{figure*}

While it is possible to classify three types of quark probability
distribution, including the $d$ quark, scalar $u$ quark and vector $u$
quark probability distributions, the scalar $u$ quark and vector $u$
quark probability distributions are not physical quantities as the two
$u$ quarks in the proton are identical particles. The proper $u$ quark
probability distribution can be obtained from the same
anti-symmetrised interpolating field of Eq.~(\ref{symInterpField}). In
spite of the symmetrisation, the $u$ quark allowed to vary prefers to
reside near the $d$ quark rather than the fixed $u$ quark as
illustrated in Fig.~\ref{LandauVCoulombUQSymd7}.

  The probability distribution of the scalar $u$ quark of
  Fig.~\ref{LandauVCoulombd6SUSurfPlot} very closely resembles that of
  the symmetrised operator, indicating that the scalar term
  contributes the most to the symmetrised probability distribution of
  Fig.~\ref{LandauVCoulombUQSymd7}

We note that there are several reasons that we are able to see diquark
clustering in the Landau gauge where Ref.~\cite{Hecht:1992uq} did
not. Our use of a large smeared source, the averaging over $\vec{x}$
in Eq.~(\ref{bigwfundef}), using improved actions for both the quarks
and the gauge fields and the consideration of hundreds of gauge fields
provides better statistics, allowing access to further $u$ quark
separations with a high signal-to-noise ratio, as well as the ability
to investigate lighter quark masses. Furthermore, our lattices extend
twice as far in the temporal direction and use fixed boundary
conditions, thus reducing the chance of any contamination associated
with the boundary conditions.

Although models featuring diquarks within hadrons have been used
extensively for many years \cite{Anselmino:1992vg}, there has been
little, if any, direct evidence for the existence of such a cluster
within a particle. Earlier lattice studies that have paired two light
quarks with a static quark \cite{Alexandrou:2005zn,Alexandrou:2006cq}
have shown a large diquark ($\mathcal{O}(1)\,\mathrm{fm}$) can form
inside of a baryon, though with limited effect on the structure of the
particle. More recently, light quarks have been paired with various
diquark correlators \cite{DeGrand:2007vu} which suggest that diquarks
are not a significant factor in light baryons. To the best of our
knowledge, this is the first time that such a diquark configuration
has been shown in a baryon composed of three light quarks.

\section{Background Fields on the Lattice} \label{BGField}
A background electromagnetic field can be added to the lattice in the
form of a phase that multiplies the $SU(3)$ links across the entire
lattice. In this case, we wish to place a constant background magnetic
field in the $z$ direction, or $\vec{B}=(0,0,B)$.  In order to
accomplish this we note that in the continuum $B_z=\del_xA_y^{EM} - \del_yA_x^{EM}$,
where $A^{EM}$ is a $U(1)$ vector potential \cite{Smit:1986fn}. As
such we need to modify this vector potential such that the magnetic
field can remain constant across the periodic boundary conditions of
the lattice. The definition of the plaquette in the $xy$ plane at some
point $x$ is given by
\eqn{W^{EM}_{\mu\nu}(x)=U^{EM}_{\mu}(x)U^{EM}_{\nu}(x+a\hat{\mu})U^{\dagger
    EM}_{\mu}(x+a\hat{\nu})U^{\dagger EM}_{\nu}(x),} where
$U^{EM}_{\mu}(x)=e^{iaeA_{\mu}(x)}$, where $a$ is the lattice spacing, and
$e$ is the electromagnetic coupling constant. Using a finite
difference approximation to the derivative, this becomes
\eqn{W^{EM}_{\mu\nu}(x)=e^{ia^2eF_{\mu\nu}(x)}.}
Using the above definition for the magnetic field strength, our focus is on
\eqn{W^{EM}(x)\equiv W^{EM}_{xy}(x)=-W^{EM}_{yx}(x)=e^{ia^2eB}.}
There are multiple vector potentials that allow such a field, two of
which will be considered here. In the first of the two, we set
$U_y(x,y,z,t)=e^{iaeBx}$ and $U_x(x,y,z,t)=1$. Away from the boundary of the
lattice, this gives
\begin{align}
W^{EM}(x,y,z,t) &= e^{iaeB(x+a)-iaeBx}\nn \\
&= e^{ia^2eB},
\end{align}
as required. On the boundary in the $x$ direction, the periodic
boundary conditions come into effect and the vector potential has to
be modified in order that the field remains constant. This is
accomplished by setting $U_x(N_x,y,z,t)=e^{-iaeN_xBy}$, where $N_x$ is the
extent of the lattice in the $x$ direction, i.e.  only on the boundary. The
plaquette then becomes
\begin{align}
W^{EM}(N_x,y,z,t)&=e^{iae(-N_xBy+Ba+N_xB(y+a)-BN_xa)}\nn \\
&=e^{ia^2eB},
\end{align}
as required. On the corner of the $xy$ plane, quantisation conditions for the field emerge
\begin{align}
W^{EM}(N_x,N_y,z,t)&=e^{ia^2B(-N_xN_y+1+N_x-N_x)}\nn \\
&=e^{ia^2eB}e^{-ia^2eN_xN_yB},
\end{align}
where $N_y$ is the extent of the lattice in the $y$ direction. Hence,
for the field to be constant at the corner of the lattice, it must be
quantised such that
\eqn{eB=\frac{2\pi n}{N_xN_ya^2},
\label{qconds} }
where $n$ is a non-zero integer.
The second method of placing a constant magnetic field on the lattice
used here is to set $U_y=1$ and $U_x=e^{-iaeBy}$ away from the
boundary and setting $U_y=e^{iaeN_yBx}$ for $x=(x,N_y,z,t)$. This implementation
has the same quantisation conditions as in Eq. (\ref{qconds}).

There are several points to note about placing a background field on
the lattice, the first of which is that adding any constant to the
potential will not affect the resultant field. It can also be shown that
there is a gauge transformation that links both of the above
implementations of the background field, given by,
\eqn{G(x,y)=e^{ieB xy},}  
where $x$, $y$ denote lattice sites $1,2,\ldots,N_x,N_y$ in units of the lattice spacing $a$ and 
\eqn{U_{\mu}(x)\rightarrow G(x)U_{\mu}(x)G^{\dagger}(x+\hat\mu).}
These implementations of the background field are applied to both
the Landau and Coulomb-fixed configurations.

We expect that this magnetic field will cause a distortion of the probability
distribution, as the proton responds to the presence of the field.
Since the magnetic field is in the $z$ direction, we
expect that physical distortion will be symmetric about this direction, and
all other effects will be a result of the choice of the gauge potential $\vec{A}$.

A particle on the lattice in the presence of a background magnetic
field will undergo a mass shift given by
\eqn{m(\vec{B})=m(0)+\frac{|e\vec{B}|}{2m}+\mu\cdot
  \vec{B}+\frac{1}{2}\beta_mB^2, \label{energyInField} } where $\mu$ is the
magnetic moment of the particle and $\beta_m$ is the magnetic
polarisability \cite{Martinelli:1982cb}. Because of the quantisation imposed by the periodic
boundary conditions, the magnetic field will be very large. For $n=3$, required to accommodate the fractional charges,
the value of the field on our lattices is $
eB=0.175~\mathrm{GeV}^2$, which implies that the first order
response of a proton to the field would be $\mu B = 260~\mathrm{MeV}$
in the continuum. On the lattice however, the mass of the ground state
of the proton is larger and the moment itself is
smaller\cite{Lee:2005ds}, and as such the response will be smaller at approximately $150~\mathrm{MeV}$ at our lighter mass.

\section{Background Magnetic Field Results} \label{BFResults}

\begin{figure*}[tph]
  \begin{center}
   $\begin{array}{c@{\hspace{0.3cm}}c}  
  \includegraphics[width=0.49\linewidth]{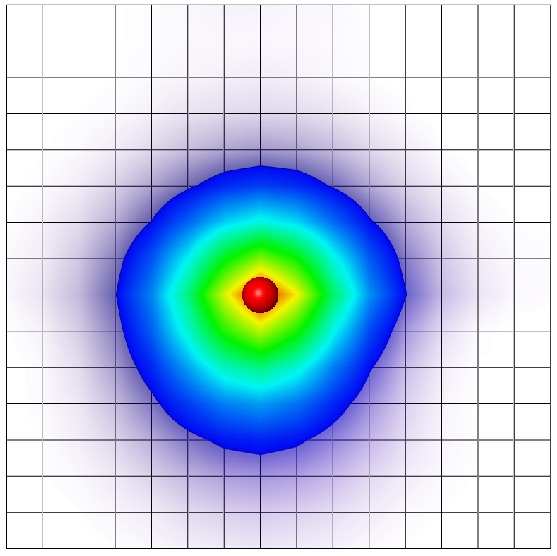} &
  \includegraphics[width=0.49\textwidth]{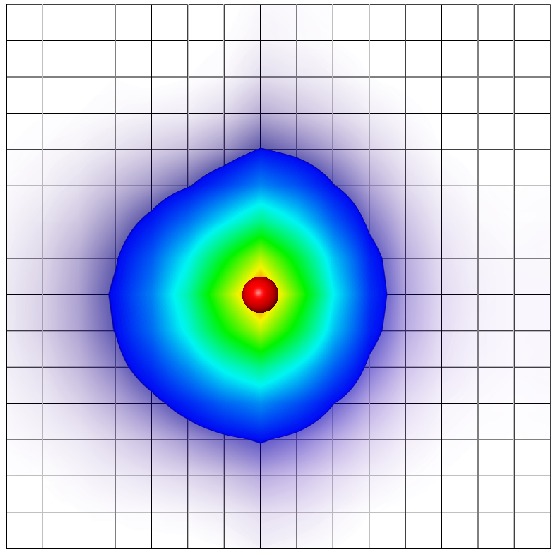}
    \end{array}$
    \caption{(Colour online) The probability distribution for the $d$ quark cut in the $x-y$ plane
      of the $u$ quarks, in the presence of a background magnetic
      field in the Landau gauge, with the first implementation (left),
      and the second implementation (right) of the vector potential
      described in Sec.~\ref{BGField}. In this image, the field,
      $\vec{B}$, is pointing into the page. The red sphere denotes the
      location of the remaining quarks. There is a clear asymmetry
      perpendicular to the field that changes with the vector
      potential, $A_{\mu}$, in spite of the background magnetic field
      not changing.}
   \label{BGFieldAsym}
  \end{center}
\end{figure*}

The first notable result from the use of the aforementioned method of
placing a background field on the lattice is that an asymmetry is
produced in the direction of the changing vector potential as
illustrated in Fig.~\ref{BGFieldAsym}. This asymmetry occurs in both
the Landau gauge and Coulomb gauge to a similar extent. This is an
unphysical result of the gauge-dependent method in which we place the
field on the lattice, which can be shown by using the second
implementation described in Sec.~\ref{BGField}. Upon doing this,
the asymmetry in the probability distribution can be seen to move to the
direction of the vector potential once again as shown in
Fig.~\ref{BGFieldAsym}. In order to minimise the effect of the choice
of the gauge potential on the probability distribution, we choose an average over
four implementations of the background field: the two implementations
described above and two in which a gauge transformation is
applied such that the magnitude of the vector potential decreases
across the lattice. For the first implementation

\eqn{G(x,y)=e^{iaeBN_xy},
\label{gtrans}}
and similarly for the second of the two implementations. Once
averaging over the four vector potentials has been applied, symmetry
around the $z$ axis is obtained. Thus, we look at the probability
distribution in the $xz$ plane.

\begin{figure*}[tph]
  \begin{center}
   $\begin{array}{c@{\hspace{0.3cm}}c}  
 \includegraphics[width=0.49\linewidth]{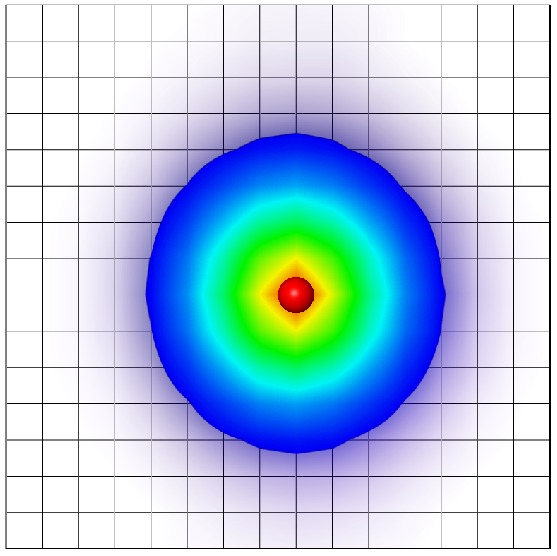} &
 \includegraphics[width=0.49\textwidth]{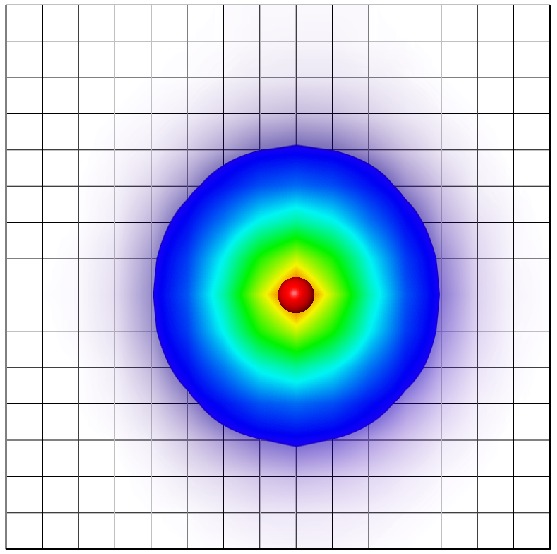}
    \end{array}$
    \caption{(Colour online) The probability distribution for the $d$ quark cut in the $x-z$ plane
      of the $u$ quarks, after symmetrising the vector potential,
      $A_{\mu}$ in the presence of the field in the Landau gauge
      (left) and Coulomb gauge (right). In this image, the field,
      $\vec{B}$, is pointing to the top of the page, and the $u$
      quarks are both in the centre of the lattice, denoted by the red sphere. In spite of the
      magnitude of the field, a fairly small deviation from spherical symmetry
      is seen in both gauges.}
   \label{BGFieldDQcomparison}
  \end{center}
\end{figure*}

In spite of the very large magnetic field strength imposed by the
boundary conditions, the change in the probability distribution is
quite small for the case where the remaining quarks are both located
in the centre of the lattice, (Fig.~\ref{BGFieldDQcomparison}). This
subtle result is consistent with that expected from the polarisablilty
as the current experimental value for the proton polarisability is
$\beta_M = 1.9(5)\times10^{-4} \, \mathrm{fm}^3$ which gives the
second order response to the field of around,
$\frac{1}{2}\beta_Me^2B^2 = 0.4\,\mathrm{MeV}$.

\begin{figure*}[tph]
  \begin{center}
   $\begin{array}{c@{\hspace{0.3cm}}c}  
 \includegraphics[width=0.49\linewidth]{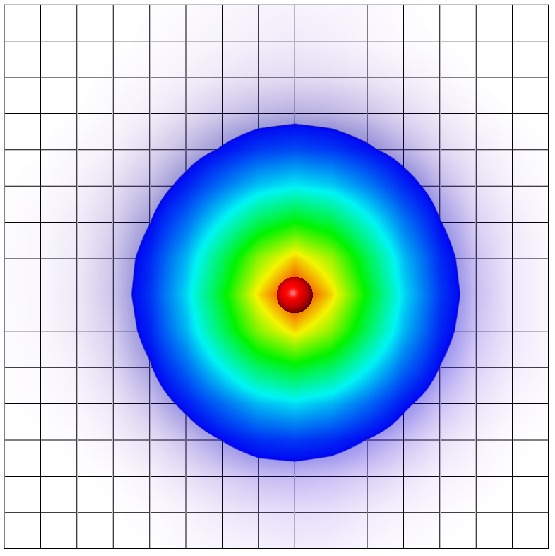} &
 \includegraphics[width=0.49\textwidth]{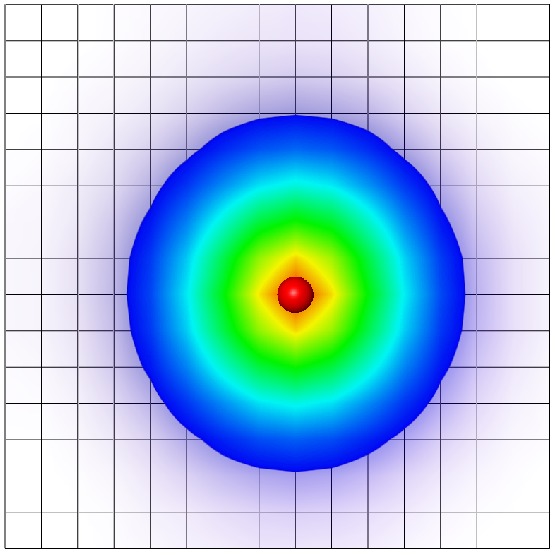}
    \end{array}$
    \caption{(Colour online) The probability distribution for the vector $u$
      quark in the presence of the background field, cut in the $x-z$
      plane of the remaining quarks in the Coulomb gauge with the spin
      aligned (left) and anti-aligned (right) to the field. The
      direction of the field is down the page, and the red sphere
      denotes the remaining quarks. The probability distribution appears more
      spherical and localized when aligned with the field, and a very
      subtle asymmetry is present in the direction of the field. The
      smallest value shown for both probability distributions is 10\% of the peak
      value.}
   \label{VUQCGspincomparison}
  \end{center}

  \begin{center}
   $\begin{array}{c@{\hspace{0.3cm}}c}  
 \includegraphics[width=0.49\linewidth]{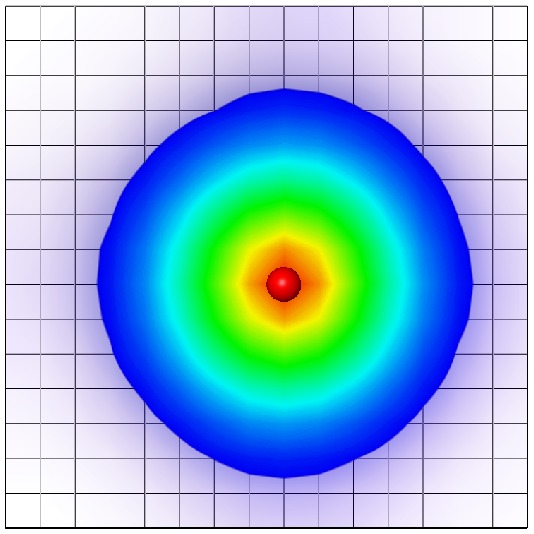} &
 \includegraphics[width=0.49\textwidth]{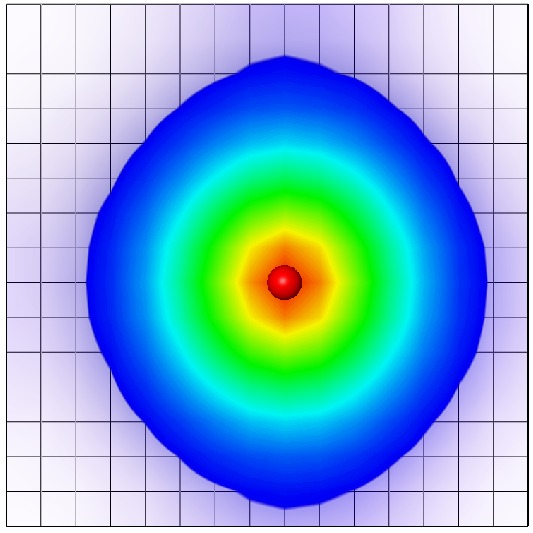}
    \end{array}$
    \caption{(Colour online) The probability distribution of the vector $u$ quark
      in the presence of the background field, cut in the $x-z$ plane
      of the remaining quarks in the Landau gauge with the spin
      aligned (left) and anti-aligned (right) to the field, and the
      red sphere denotes the remaining quarks. The direction of the
      field is down the page. Much like in the Coulomb gauge, the probability
      distribution appears more spherical and localized when aligned with
      the field. The smallest value shown for both probability distributions is
      10\% of the peak value.}
   \label{VUQLGspincomparison}
  \end{center}
\end{figure*}

Very little spin dependence can be seen in the probability distributions
themselves, the probability distributions of the spin up proton quarks are
largely the same as the probability distributions of the spin down proton. A
subtle difference appears in the vector $u$ quark probability distributions in
the Coulomb gauge, as illustrated in Fig.~\ref{VUQCGspincomparison}.
A more prominent difference is visible in the Landau gauge (Fig.~\ref{VUQLGspincomparison}). 
The probability distribution appears more spherical
and localized when the spin is aligned with the field, and a very
subtle asymmetry is present in the direction of the field. Spin
dependence also manifests itself in the energy of the proton, as can
be seen in Table \ref{masstable}, where the energy of the proton when
its spin is anti-aligned to the field is lower than the zero-field
energy, indicating that Landau levels are not having a dominant
effect on the particle energy. The spin aligned proton receives a
larger energy, due to the sign on the moment term.

\begin{table}[h]
\caption{The dependence of the spin up and spin down mass of the
  proton on the background magnetic field. When the spin is aligned
  with the field (up), the mass of the proton increases, whereas when the spin
  is anti-aligned with the field (down), we see a mass decrease.}
\label{masstable}
\begin{tabular}{cccccccc}
\hline\hline
$\kappa$ & spin & $B$ & Mass (GeV) & $m_\pi^2$ ($\mathrm{GeV}^2$) & window & $\chi^2/\mathit{dof}$ \\
\hline
0.12885  & averaged & 0  & 1.492(10) & 0.486 &  10-18  &  1.001  \\
         & down & -3 & 1.366(11) &       &  10-14  &  0.879  \\
         & up   & -3 & 1.688(11) &       &  10-18  &  0.991  \\
0.12990  & averaged & 0  & 1.327(11) & 0.283 &  10-18  &  0.954  \\
         & down & -3 & 1.197(13) &       &  10-14  &  1.061  \\
         & up   & -3 & 1.528(13) &       &  10-15  &  0.983  \\

\hline\hline
\end{tabular}
\end{table}

The localization of the spin aligned probability distribution can be understood
in terms of a constituent quark mass effect in a simple potential
model. The effect of the increased proton energy is to cause
an increase in the constituent quark mass, hence causing the probability
distribution to sit lower in the potential. This makes the spin aligned
probability distribution smaller than the spin anti-aligned probability distribution.

\begin{figure*}[tph]
  \begin{center}
    $\begin{array}{c@{\hspace{0.3cm}}c}  
     \includegraphics[width=0.49\linewidth]{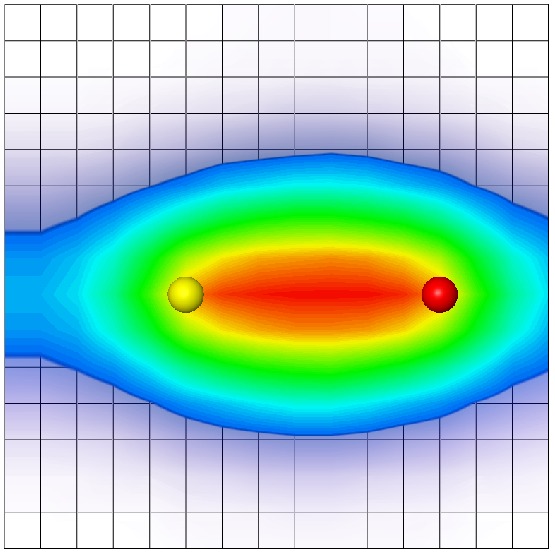} &
     \includegraphics[width=0.49\textwidth]{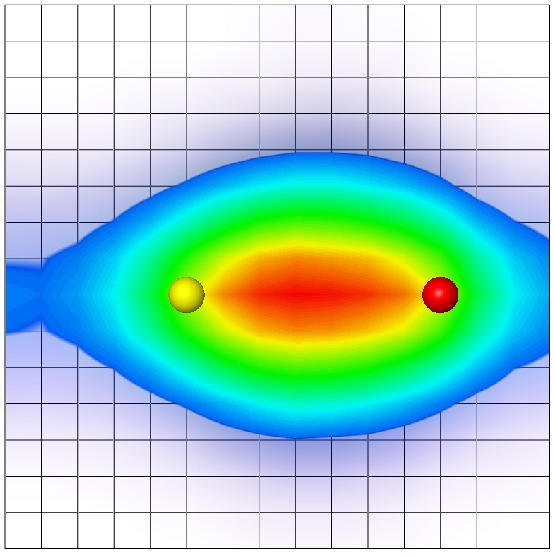}
    \end{array}$
    \caption{(Colour online) The probability distribution of the $d$
      quark in the Coulomb gauge cut in the $x-z$ plane of the $u$
      quarks which are separated by seven lattice units in the
      transverse direction with zero background field (left) and in
      the presence of the field (right). The direction of the field is
      up the page and the spheres denote the positions of the $u$
      quarks. The smallest value shown for both probability
      distributions is 20\% of the peak value.}
   \label{CGlargeQsepComparison}
  \end{center}

\begin{center}
   $\begin{array}{c@{\hspace{0.3cm}}c}  
 \includegraphics[width=0.49\linewidth]{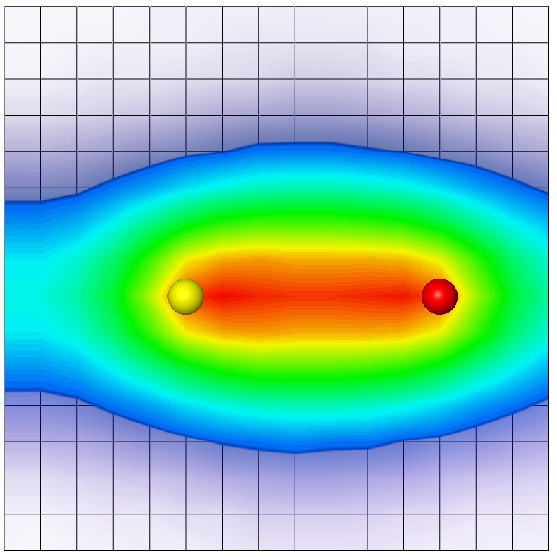} &
 \includegraphics[width=0.49\textwidth]{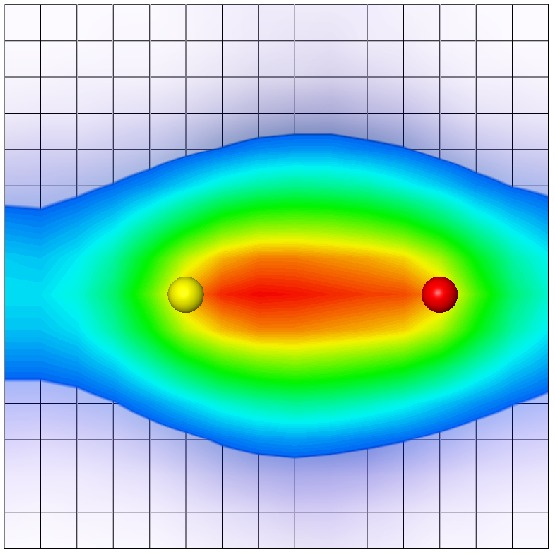}
    \end{array}$
    \caption{(Colour online) The probability distribution of the $d$ quark, in
      the Landau gauge cut in the $x-z$ plane of the remaining quarks
      which are separated by 7 lattice units in the transverse
      direction with zero background field (left) and in the presence
      of the field (right). The spheres denote the positions of the
      $u$ quarks. The diquark clustering is barely visible in this view,
      and disappears completely in the presence of the field. The probability
      distributions are broader in the Landau gauge and the smallest value
      shown for both probability distributions is 20\% of the peak value.}
   \label{LGlargeQsepComparison}
  \end{center}
\end{figure*}

As the quarks are separated, the probability distributions in the
background field tend to be more localized than the same probability
distributions without a background field. Some stretching along the
field orientation at the centre of the distribution is apparent,
making the distribution more spherical
(Fig.~\ref{CGlargeQsepComparison}). This is consistent with the
effect of raising the constituent quark mass. In the Landau gauge, the
diquark clustering is removed from the $d$ quark probability
distribution by the presence of the field as illustrated in
Fig.~\ref{LGlargeQsepComparison}.

\begin{figure*}[tph]
  \begin{center}
    $\begin{array}{c@{\hspace{0.3cm}}c}  
     \includegraphics[width=0.49\linewidth]{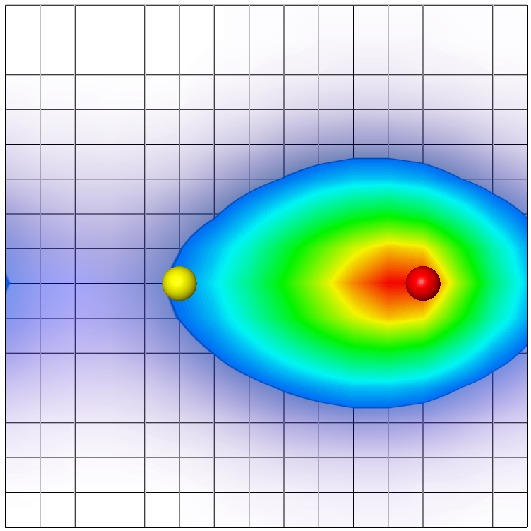} &
     \includegraphics[width=0.49\textwidth]{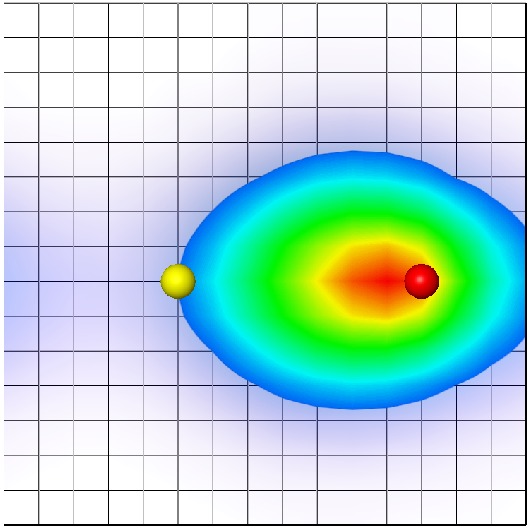}
    \end{array}$
    \caption{(Colour online) The probability distribution of the scalar $u$ quark
      in the Coulomb gauge cut in the $x-z$ plane of the remaining
      quarks which are separated by seven lattice units in the
      transverse direction with zero background field (left) and in
      the presence of the field (right). The direction of the field is
      up the page and the $d$ quark is on the right, denoted by the
      red sphere. In contrast to the $d$ quark probability distribution, there is
      still a distinct preference for the formation of a scalar
      diquark. When the field is applied, the probability distribution can be
      seen to move toward the centre of the lattice. The smallest
      value shown for both probability distributions is 20\% of the peak value.}
   \label{CGSUQlargeQsepComparison}
  \end{center}

\begin{center}
   $\begin{array}{c@{\hspace{0.3cm}}c}  
 \includegraphics[width=0.49\linewidth]{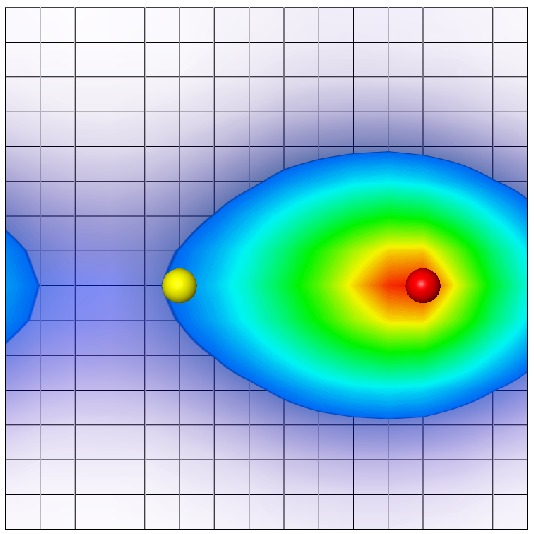} &
 \includegraphics[width=0.49\textwidth]{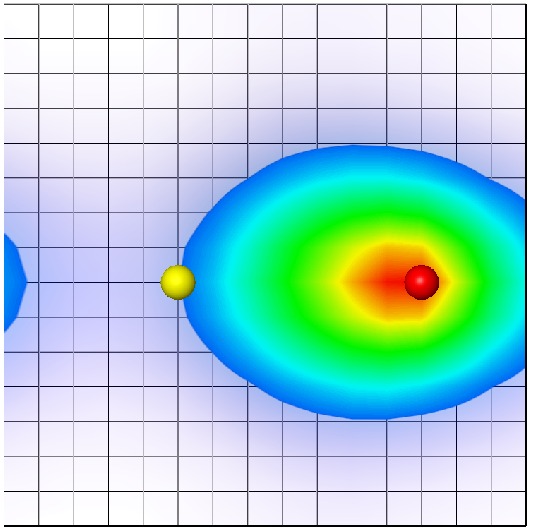}
    \end{array}$
    \caption{(Colour online) The probability distribution of the scalar $u$
      quark, in the Landau gauge which are separated by 7 lattice
      units in the transverse direction with zero background field
      (left) and in the presence of the field (right).  The direction
      of the field is up the page and the $d$ quark is on the
      right, denoted by the red sphere. Preference towards the centre of the lattice is also
      visible in the Landau gauge, but is more subtle than in the
      Coulomb gauge. The probability distributions are broader in the Landau
      gauge and the smallest value shown for both probability distributions is
      20\% of the peak value.}
   \label{LGSUQlargeQsepComparison}
  \end{center}
\end{figure*}

\begin{figure*}[tph]
  \begin{center}
    $\begin{array}{c@{\hspace{0.3cm}}c}  
     \includegraphics[width=0.49\linewidth]{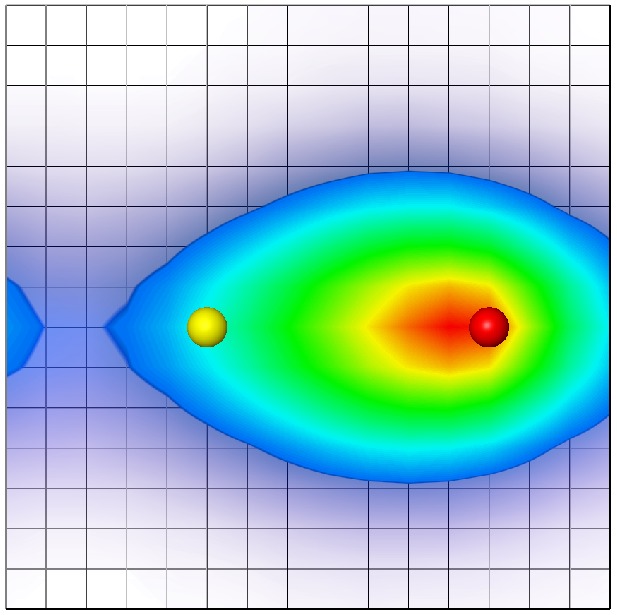} &
     \includegraphics[width=0.49\textwidth]{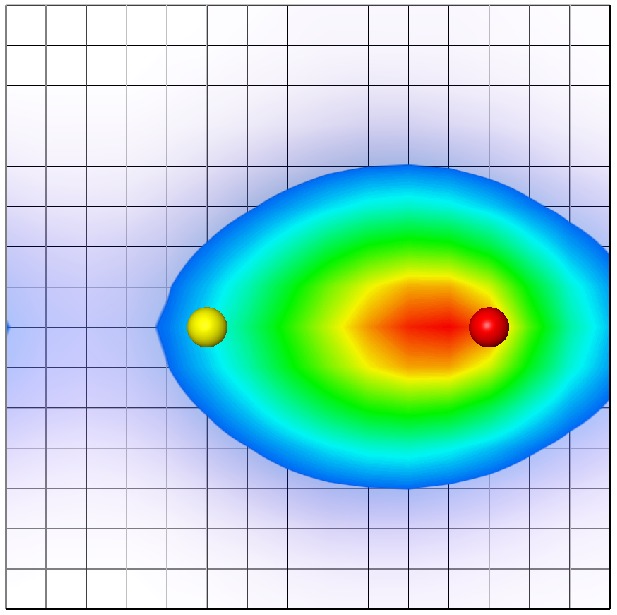}
    \end{array}$
    \caption{  (Colour online) The probability distribution of a $u$
      quark in the Coulomb gauge cut in the $x-z$ plane of the
      remaining quarks which are separated by seven lattice units in
      the transverse direction with zero background field (left) and
      in the presence of the field (right). The direction of the field
      is up the page and the $d$ quark is on the right, denoted by the
      red sphere. The symmetrised $u$ quark probability distribution
      bears close resemblance to the scalar $u$ quark, but less
      is localized due to the vector $u$ quark contribution. The smallest
      value shown for both probability distributions is 20\% of the
      peak value.} 
   \label{CGSymUQlargeQsepComparison}
  \end{center}

\begin{center}
   $\begin{array}{c@{\hspace{0.3cm}}c}  
 \includegraphics[width=0.49\linewidth]{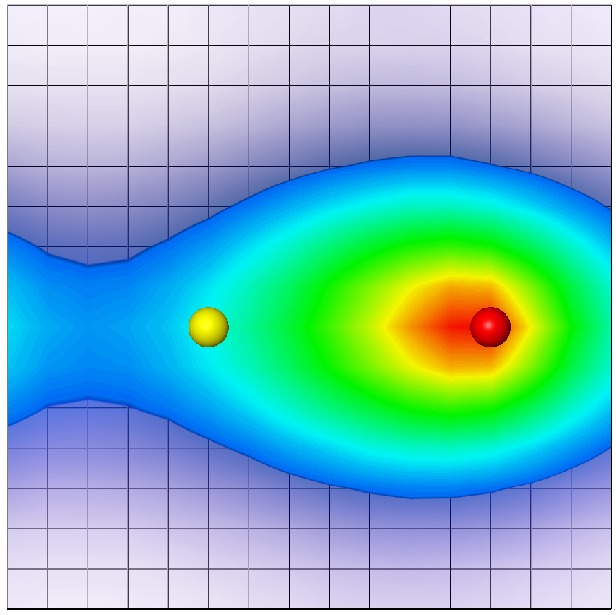} &
 \includegraphics[width=0.49\textwidth]{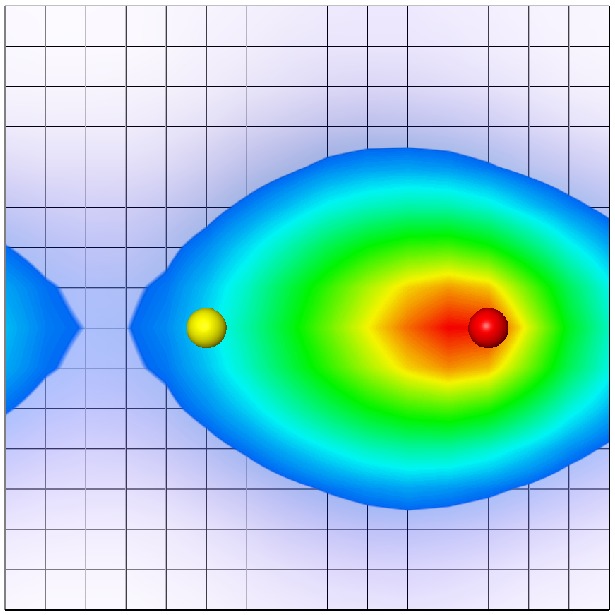}
    \end{array}$
    \caption{ (Colour online) The probability distribution
        of a $u$ quark, in the Landau gauge cut in the $x-z$ plane of
        the remaining quarks which are separated by 7 lattice units in
        the transverse direction with zero background field (left) and
        in the presence of the field (right).  The direction of the
        field is up the page and the $d$ quark is on the right,
        denoted by the red sphere. The contribution to the symmetrised
        probability distribution from the vector $u$ quark is enhanced
        in the Landau gauge compared to the Coulomb gauge. The
        smallest value shown for both probability distributions is
        20\% of the peak value.} 
   \label{LGSymUQlargeQsepComparison}
  \end{center}
\end{figure*}

 In contrast, diquark clustering is still apparent in the $u$ quark
 probability distribution in the presence of the field, with the
 distribution moving towards the centre of the baryon on application
 of the magnetic field, as shown in
 Figs.~\ref{CGSUQlargeQsepComparison} and
 \ref{LGSUQlargeQsepComparison}. The scalar $u$ quark probability
 distribution also shows more localization than either the vector $u$
 quark or $d$ quark probability distributions. The anti-symmetrised
 $u$ quark probability distribution illustrated in
 Figs.~\ref{CGSymUQlargeQsepComparison} and
 \ref{LGSymUQlargeQsepComparison} still bears close resemblance to
 that of the scalar $u$ quark. However, it is not as localized as the
 scalar $u$ quark probability distribution due to the contribution
 from the vector $u$ quark required to anti-symmetrise the identical
 $u$ quarks. The Landau gauge probability distribution is still larger
 than the Coulomb gauge probability distribution.

\begin{figure*}[tph]
  \begin{center}
    $\begin{array}{c@{\hspace{0.3cm}}c}  
     \includegraphics[width=0.49\linewidth]{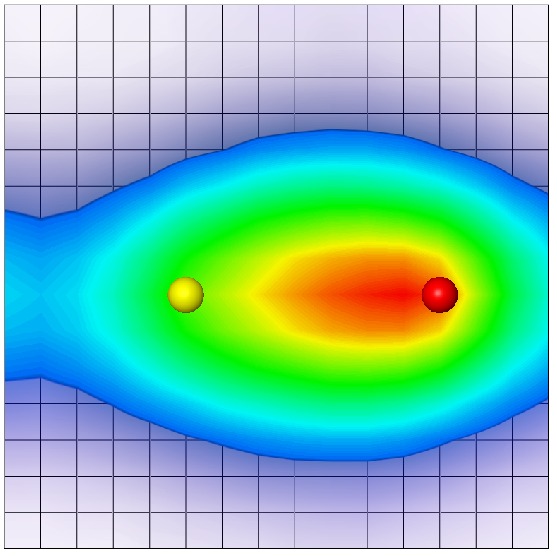} &
     \includegraphics[width=0.49\textwidth]{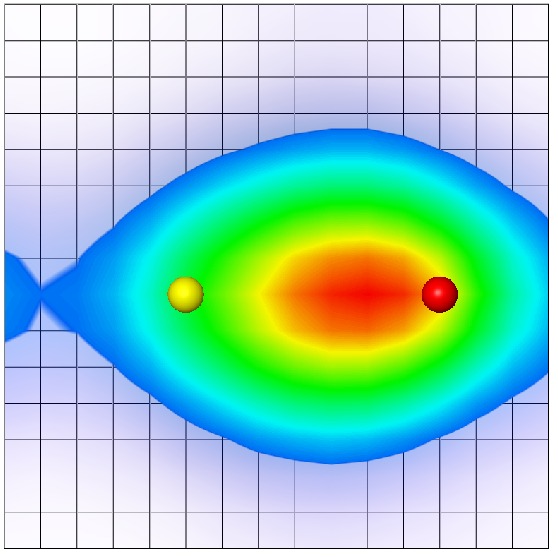}
    \end{array}$
    \caption{(Colour online) The probability distribution of the vector $u$ quark
      in the Coulomb gauge cut in the $x-z$ plane of the remaining
      quarks which are separated by seven lattice units in the
      transverse direction with zero background field (left) and in
      the presence of the field (right). The direction of the field is
      up the page and the $d$ quark is on the right, denoted by the
      red sphere. The effect of the field on the vector $u$ quark probability
      distribution is more pronounced than the $d$ quark and scalar $u$
      quark probability distributions. The smallest value shown for both probability
      distributions is 20\% of the peak value.}
   \label{CGVUQlargeQsepComparison}
  \end{center}

\begin{center}
   $\begin{array}{c@{\hspace{0.3cm}}c}  
 \includegraphics[width=0.49\linewidth]{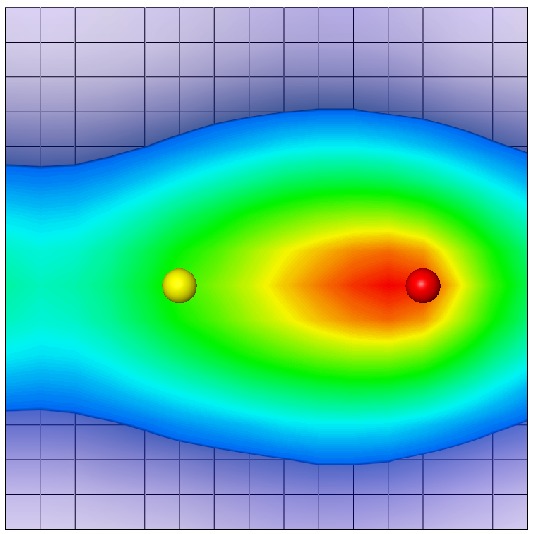} &
 \includegraphics[width=0.49\textwidth]{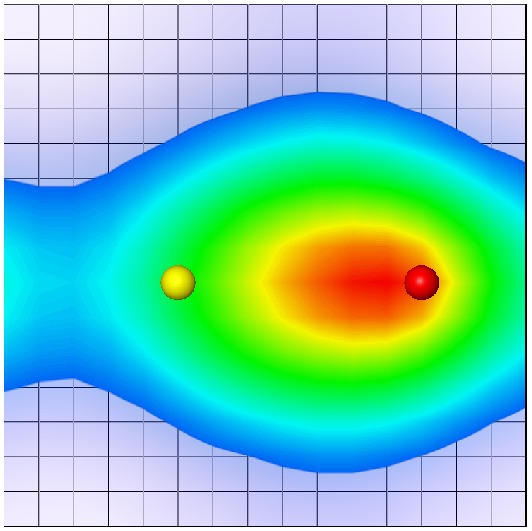}
    \end{array}$
    \caption{(Colour online) The probability distribution of the vector $u$
      quark, in the Landau gauge cut in the $x-z$ plane of the
      remaining quarks which are separated by 7 lattice units in the
      transverse direction with zero background field (left) and in
      the presence of the field (right). The direction of the field is
      up the page and the $d$ quark is on the right, denoted by the
      red sphere. The smallest value shown for both probability distributions is
      20\% of the peak value.}
   \label{LGVUQlargeQsepComparison}
  \end{center}
\end{figure*}

As illustrated in Figs.~\ref{CGVUQlargeQsepComparison} and
\ref{LGVUQlargeQsepComparison} for the Coulomb and Landau gauges
respectively, the effect of the field on the probability distribution
of the vector $u$ quark is more pronounced than the $d$ quark and
scalar $u$ quark probability distributions.

The spin orientation dependence as the quarks are separated remains
largely the same as in the case where the quarks are at the origin,
with the vector $u$ quark probability distribution changing the most
between the spin aligned and anti-aligned cases. In the case where the
spin is aligned with the field and the mass increases, the probability
distribution becomes more localized perpendicular to the field
relative to when the spin is anti-aligned with the field. This is in
keeping with the constituent quark model, where the field causes the
constituent quark mass to increase, and as such, the proton sits lower in the
potential.

 Very little spin dependence is visible in the $d$ quark and scalar
 $u$ quark probability distributions. However, the effect on the probability
 distribution due to the magnetic field is more prominent when
 the remaining quarks are separated, compared to when the quarks are
 at the origin.


\section{Conclusion} \label{Conclusion}

In this study, we have performed the first examination of the probability
distribution of quarks in the proton in the presence of a background magnetic field
in both the Landau and Coulomb gauges. 

We have shown that there is a distinct difference between the $d$
quark probability distributions in the Landau and Coulomb gauge, with
the Landau gauge exhibiting clear diquark clustering. The probability
distributions in the Coulomb gauge did not. The scalar $u$ quark and
vector $u$ quark probability distributions show clear diquark
clustering in both the Landau and Coulomb gauge, with the scalar $u$
quark being more tightly bound to the $d$ quark than the vector $u$
quark probability distribution. This is the first direct evidence of
the ability of a scalar diquark pair to form in a baryon. Also, the
probability distributions in the Landau gauge were larger than those
in the Coulomb gauge.

On the application of the background field, we found a gauge
dependence in the probability distribution in the direction of the
vector potential. A symmetrisation was performed to rectify this. 

In spite of the very large magnetic field required by the quantisation
conditions, the change in the probability distribution is small, being
most prominent in the vector $u$ quark. The effect is to elongate the
distribution along the axis of the field while generally localizing
the distribution. The vector $u$ quark exhibits the most spin
dependence, with the probability distribution being more localized
when the spin is aligned with the magnetic field. This effect can be
understood in terms of a constituent quark model where the constituent
quark mass increases in the presence of the magnetic field.

More notable spin dependence appeared in the energy of the proton
itself, largely associated with the magnetic moment, as opposed to
higher order effects impacting the structure of the proton.  As the
nucleon is rather stiff and only slightly more localized in a magnetic
field, we anticipate the background field approach to determining the
magnetic moment of baryons to be effective, even in a strong
background field.

\section*{Acknowledgments}

This research was undertaken on the NCI National Facility in Canberra,
Australia, which is supported by the Australian Commonwealth
Government. We also thank eResearch SA for generous grants of
supercomputing time which have enabled this project. This research is
supported by the Australian Research Council.



\end{document}